\documentclass[12pt,a4paper]{article}
\usepackage[latin1]{inputenc}
\usepackage{amsmath}
\usepackage{amsmath} 
\usepackage{amsfonts}
\usepackage{amssymb}
\usepackage{float}
\usepackage{cite}
\usepackage{breqn}
\usepackage{graphics}
\usepackage{mathtools}
\usepackage{xcolor}
\usepackage{mathrsfs}
\newcommand{\plainfootnote}[1]{%
  \begingroup
  \renewcommand{\thefootnote}{}
  \footnotetext{#1}%
  \addtocounter{footnote}{-1}
  \endgroup
}
\usepackage{fontawesome}
\definecolor{lightred}{rgb}{1, 0.5, 0.5}
\usepackage{tikz}
\usepackage{adjustbox}
\DeclareGraphicsExtensions{.pdf,.png,.jpg}
\usepackage[colorlinks=true,linkcolor=blue,citecolor=red, urlcolor=cyan]{hyperref}
\makeatletter
\newcommand{\mathleft}{\@fleqntrue\@mathmargin0pt}
\newcommand{\mathcenter}{\@fleqnfalse}
\makeatother
\newcommand*{\SavedEqref}{}
\let\SavedEqref\eqref
\renewcommand*{\eqref}[1]{%
  \begingroup
    \hypersetup{
      linkcolor=linkequation,
      linkbordercolor=linkequation,
    }%
    \SavedEqref{#1}%
  \endgroup
}

\usepackage[a4paper, width = 170mm, top = 30mm, bottom = 25mm, 
bindingoffset = 0mm]{geometry}

\def\beq{\begin{equation}}
\def\eeq{\end{equation}}
\def\bea{\begin{eqnarray}}
\def\eea{\end{eqnarray}}

\definecolor{lightorange}{RGB}{255, 178, 102} 
\definecolor{mediumlightblue}{RGB}{100, 170, 255}   
\definecolor{mediumlightred}{RGB}{230, 80, 80}      

\begin{document}

 
\begin{center}
	{\large \bf Holonomically constrained polarization transformation
}
\vspace{0.7cm}

{\sf \small Mohammad Umar\textsuperscript{\textcolor{red}{\textnormal{*}}}, P. Senthilkumaran\textsuperscript{\textcolor{blue}{\textnormal{*}}}}

\bigskip
\plainfootnote{\textcolor{red}{\faEnvelope}\, aliphysics110@gmail.com, opz238433@opc.iitd.ac.in}

\plainfootnote{\textcolor{blue}{\faEnvelope}\textsuperscript{\textcolor{blue}{\textnormal{}}} psenthilk@yahoo.com, psenthil@opc.iitd.ac.in}

{\em
\textsuperscript{\textcolor{red}{\textnormal{}}}Optics and Photonics Centre\\  
Indian Institute of Technology Delhi\\ 
New Delhi 110016, INDIA\\}
\vspace{0.5em}


\vspace{1.2cm}	
\noindent {\bf Abstract}		
\end{center}

\noindent
In polarization optics, various topological constructs, namely Poincar\'e spheres of different orders, are used to represent uniform and structured polarization distributions. Similarly, there are also structured polarization optical elements. Consequently, various topological indices are defined for structured beams and elements.  These topological aspects naturally allow us to holonomy-based categorization of polarization transformations. In this paper, we introduce holonomically constrained polarization transformations on topological constructs. The conditions on the topological parameters of the beams, elements and spheres to achieve holonomically constrained polarization transformations are discussed in detail. A topological treatment of holonomic systems is needed, since abundant polarization transformations reported in the literature on beams with structured polarization are nonholonomic. It is prudent to carry out this study since it can not be assumed that switching between holonomic and nonholonomic transformations does not affect the system output. It is shown here how these concepts enable us to introduce topological index spaces for polarization optics.

\medskip
\vspace{1in}
\newpage
	

\section{Introduction}
Constraints limit the motion of a system; for example, pearls in a necklace are constrained to move along the curve defined by the supporting string \cite{goldstein2011classical}. Constraints also exist in optics. The holonomic constraint in polarization optics is defined by the constraint equation ${S_1}^2+{S_2}^2+{S_3}^2={S_0}^2$ for a fully polarized light, where $S_0$, $S_1$, $S_2$ and $S_3$ are the standard Stokes parameters (SPs) defining the state of polarization (SOP) of light \cite{goldstein2003polarized}.  For a partially polarized light the constraint equation is modified to ${S_1}^2+{S_2}^2+{S_3}^2<{S_0}^2$ - similar to gas molecules in a closed container are constrained to move within the container. In this paper, we focus on fully polarized light. Any polarization transformation between states A and B can be mapped to the surface of the Poincar\'e sphere (PS) by a trajectory connecting the two states.  Every \textit{state} and the \textit{trajectory} are constrained to be on the surface of the PS. Therefore, every polarization transformation using retarders is holonomic on the standard PS and this is a canonical example of holonomic polarization transformations. \\
\indent
Non-holonomic transformations become possible with the advent of structured light featuring polarization singularities \cite{ruchi2020phase}, which can be mapped as points on various topological constructs \cite{padgett1999poincare, milione2011higher, yi2015hybrid, arora2020hybrid}. A vector field singularity \cite{zhan2009cylindrical}, represented by a point on a higher-order Poincar\'e sphere (HOPS) \cite{milione2011higher}, can no longer be described by a point on the same HOPS after passing through a retarder plate. A suitable example of this type of transformation is the polarization singularity index inversion by using a half-wave plate (HWP) \cite{pal2017polarization}, thereby rendering the polarization transformation non-holonomic. \\
\indent
In recent times, there has been great interest in employing (spin and orbital) angular momentum states of light in communication \cite{yan2014high, wang2012terabit, yu2015potentials}, where beams with different topological indices serve as different channels wherein cross-talk between orthogonal states is absent, if not, minimal.  Polarization singularities are orthogonal superpositions of phase singularities \cite{ruchi2020phase}. In the channel, if at all any polarization transformation occurs, a holonomically constrained transformation is preferred. In a non-holonomic transformation, the beam loses its topological signature. However, holonomic polarization transformations are possible in a topological construct such as a HOPS, where the states and trajectories are constrained to the surface of the respective spheres. We also notice that in holonomic polarization transformations, the topological index of the beam is conserved means the beam does not looses its topological signature. Holonomic systems allow for the introduction of the concept of topological index space, which corresponds to the topological sphere and polarization circuits, offering deeper insights into polarization optics\\
\indent
In this article, we discuss the conditions required to realize holonomically constrained polarization transformations on topological constructs such as the HOPS. We also introduce the concept of polarization optics with a non-zero topological index space by presenting some polarization circuits. The HOPS beam and structured optical elements possess certain topological features; hence, before venturing into these novel ideas, we first outline the essential background on the topological parameters of beams, topological spheres and polarizing elements.\\
\indent
The structure of the paper is as follows: Section \ref{section2} presents a discussion on the HOPS, followed by Section \ref{section3}, which addresses the concept of $q$-plate. Section \ref{section4} provides a mathematical derivation of the holonomic condition necessary for traversing the HOPS holonomically. Section \ref{section5} introduces the concept of non-holonomic transformations in polarization optics. In Section \ref{section6}, the concept of topological index space is introduced and Section \ref{section7} explores the topological index spaces for the higher-order topological spheres with their corresponding structured elements. Finally, the paper reaches its conclusion in Section \ref{section8}.
\section{The (higher-order) Poincar\'e sphere}
\label{section2}
Although there are multiple parameters to describe the state of polarization of light, two parameters, namely the azimuth $\gamma = 0.5 \tan^{-1}(S_2/S_1)$ and the ellipticity $\chi =0.5 \sin^{-1}(S_3/S_0)$ of the polarization ellipse, suffice \cite{hajnal1987singularities,  freund2002polarization, dennis2002polarization, zhan2009cylindrical}. These two parameters are related to the coordinates (longitude $2\gamma$ and latitude $2\chi$) on the topological sphere \cite{bansal2023stokes, bansal2023experimental}. For homogeneously polarized light, $\gamma$ and $\chi$ are constants, and the topological index is zero. A standard PS can be used to represent all conceivable polarization states. On this sphere, the poles correspond to the eigen-polarization basis states: the north pole represents right circularly polarized light, while the south pole represents left circularly polarized light. All other points on the sphere represent superposition states. For inhomogeneously polarized light fields with polarization singularities, $\gamma$ and $\chi$ vary spatially. A vector field singularity is characterized by a topological parameter known as the Poincar\'{e}-Hopf (PH) index, defined as 
\begin{equation}
    \eta = \frac{1}{2\pi} \oint \nabla \gamma \cdot dl,
\end{equation}
where $\gamma$ represents the azimuth/orientation of the polarization elliptically/linearly polarized light. Like homogeneously polarized beams, these vector vortex beams can also be represented as superposition states. Similar to the standard PS, the HOPS enables these beams to be represented as points on its surface. These types of beams are bijectively mapped onto the HOPS. In this letter, we restrict our discussion to the HOPS alone.\\
\indent
In HOPS, the basis states are the orthogonal spin angular momentum and orthogonal orbital angular momentum states \cite{yi2015hybrid, milione2011higher, arora2020hybrid} and is given by
\begin{align*}
|R_\ell\rangle = e^{-i\ell\phi}|R\rangle, \quad \quad |L_\ell\rangle = e^{i\ell\phi}|L\rangle,
\end{align*}
where $|R\rangle$ and $|L\rangle$ are the unit vectors in the circular basis and $\pm \ell$ is the topological charge of the phase vortex beams that form the orbital angular momentum basis.  Any structured vector vortex beam can be represented as a superposition state
\begin{equation}
|\psi_\ell\rangle = \psi_R^\ell |R_\ell\rangle+\psi_L^\ell |L_\ell\rangle,
\label{beam}
\end{equation}
where $\psi_R^\ell=\langle R_\ell | \psi_\ell\rangle$ and $\psi_L^\ell=\langle L_\ell | \psi_\ell\rangle$ is the $|R_\ell\rangle$ and $|L_\ell\rangle$ component of $|\psi_\ell\rangle$ respectively. For the HOPS beams given by Eq. (\ref{beam}), the Poincar\'e Hopf (PH) index is given by $\eta=\ell$. In the context of HOPS, the normalized higher-order Stokes parameters (SPs) are defined by
\begin{eqnarray}
\label{HOStokes}
S_0^{(\eta)}&=& |\psi_R^\ell|^2+|\psi_L^\ell|^2, \nonumber \\
S_1^{(\eta)}&=& 2|\psi_R^\ell||\psi_L^\ell| \cos\phi_{12}, \nonumber \\
S_2^{(\eta)}&=& 2|\psi_R^\ell||\psi_L^\ell| \sin\phi_{12}, \nonumber \\
S_3^{(\eta)}&=& |\psi_R^\ell|^2-|\psi_L^\ell|^2, 
\end{eqnarray}
where $\phi_{12}=\arg\left[S_1^{(\eta)}+iS_2^{(\eta)}\right]$ is referred as $\phi_{12}$ Stokes phase defined with higher-order SPs. These SPs are used in the construction of HOPS. In terms of these SPs, the longitude $2\gamma^{(\eta)}$ and latitude $2\chi^{(\eta)}$ are defined as
\begin{equation}
2\gamma^{(\eta)} = \tan^{-1}\left(\frac{S_{2}^{(\eta)}}{S_{1}^{(\eta)}}\right), \quad \quad 2\chi^{(\eta)} = \sin^{-1}\left(\frac{S_{3}^{(\eta)}}{S_{0}^{(\eta)}}\right).
\end{equation}
\begin{figure}[t]
\centering
\includegraphics[width=0.85\linewidth]{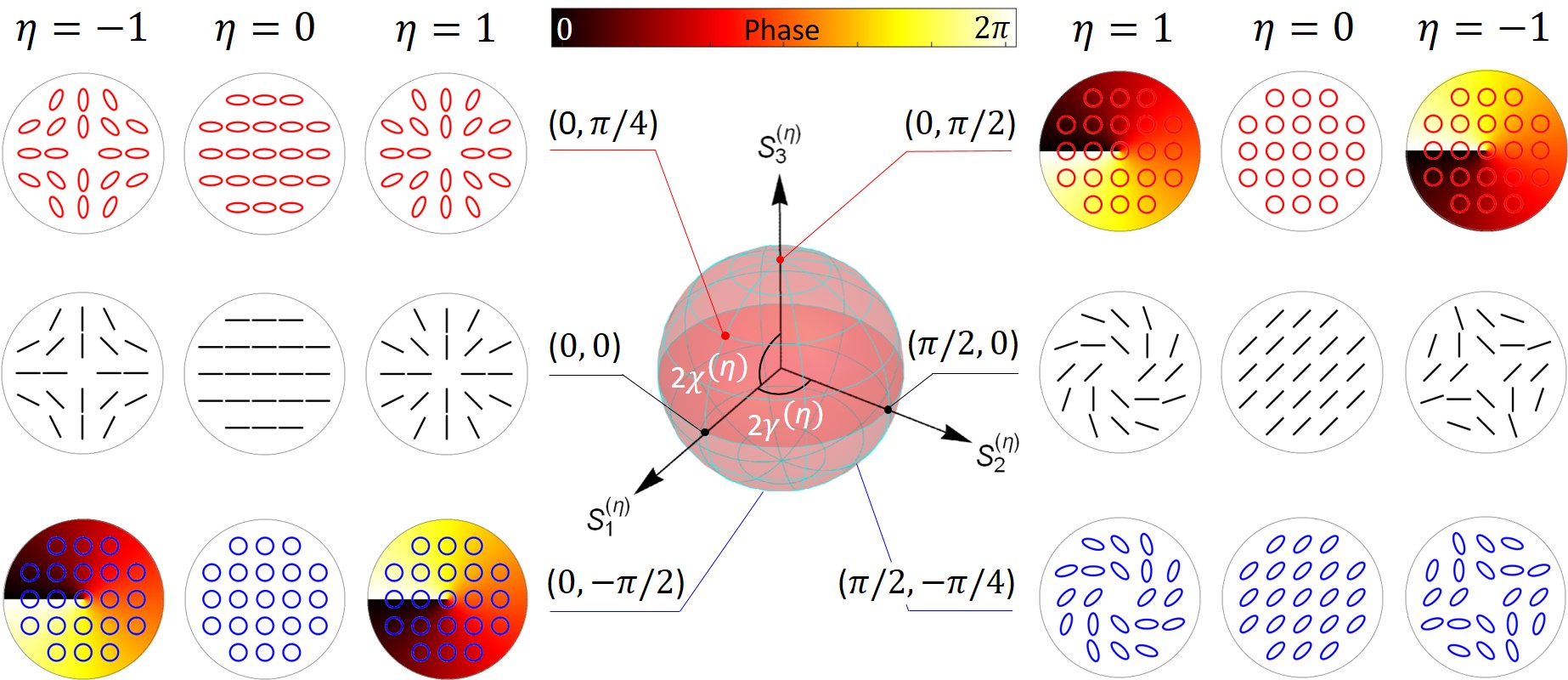}
\caption{(Color online). \textit{Polarization distribution for the topological sphere of orders $\eta = 1, 0$ and $-1$. The red and blue colors in the polarization ellipses represent right- and left-handed polarization, respectively. The coordinates on the sphere are given by ($2\gamma^{(\eta)}$, $2\chi^{(\eta)}$).  The phase distribution of the eigen states are provided as background for the two orthogonal polarization basis states.}}
\label{q01}
\end{figure}
\indent
Since there exists an infinite set of orthogonal angular momentum states, an infinite number of HOPS can be defined, each characterized by a distinct topological index $\eta$. Fig. \ref{q01} illustrates HOPS of different orders ($\eta = 0$, $1$, and $-1$), along with the corresponding polarization distributions represented as points on their respective spheres.
\section {The \textit{q}-plate: an inhomogeneous waveplate}
\label{section3}
\begin{figure}[t]
\centering
\includegraphics[width=0.43\linewidth]{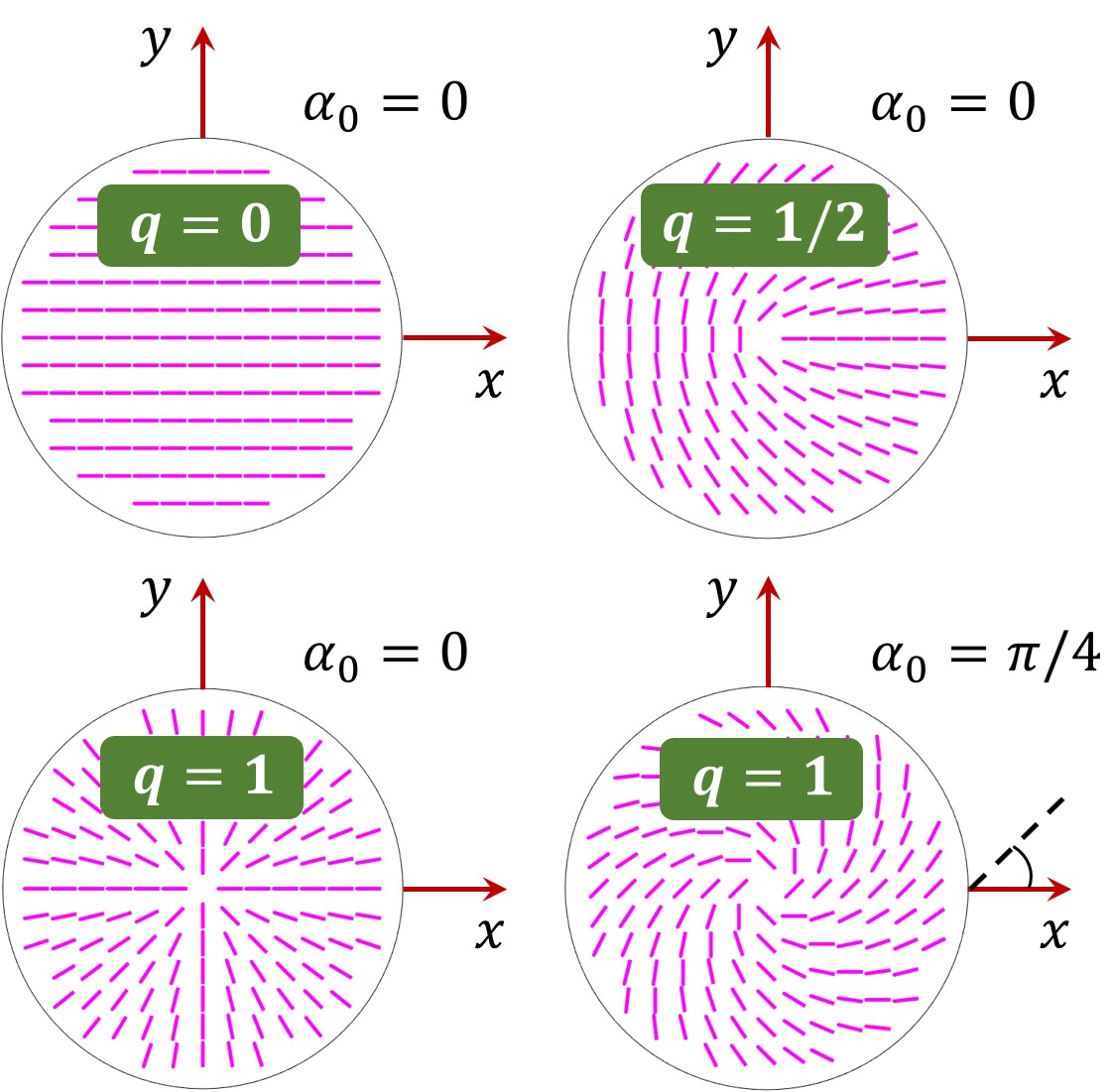}
\caption{(Color online). \textit{The geometry of the $q$-plates with $q=0$, $q=1/2$ and $q=1$ are shown. These are retarders in which the fast axis is spatially varying according to the $q$-value. For $q=0$, the retarder is a homogeneous waveplate. In the bottom row, the role of off-set angle $\alpha_0$ is depicted for the $q=1$ plate.}}
\label{q02}
\end{figure}

In recent years, inhomogeneous polarizing elements have been made of retarders, whose fast axis orientation is position-dependent.  Such plates are called S-wave plates or spatially varying retarders or $q$-plates \cite{marrucci2006optical, machavariani2008spatially, marrucci2013q, rubano2019q, gregg2015q}. These waveplates are characterized by the $q$-value (topological charge), defined by 
\begin{equation}
  q =\frac{1}{2\pi} \oint \nabla \alpha(\phi) \cdot dl.  
\end{equation} 
where, $\alpha(\phi)$ is the fast axis orientation of the $q$-plate and defined by $\alpha(\phi) = q\phi + \alpha_{0}$. Here, $q$ is the topological charge that quantifies the rotation of the fast axis over a full $2\pi$ cycle, and $\alpha_{0}$ is the offset angle relative to a reference axis. Based on this definition, $q$-plates with a specific topological charge $q$ encompass all retarders with retardance $\delta$. Throughout this paper, we use the notation for $q$-plate as $q^{\delta}$. A $q$-plate with a retardance of $\pi$ is referred to as a half-wave $q$-plate ($q^{H}$-plate or $q$-HWP), while one with a retardance of $\pi/2$ is called a quarter-wave $q$-plate ($q^{Q}$-plate or $q$-QWP). Fig. \ref{q02} shows the fast axis orientation distributions of $q$-plates for various $q$-values. The Jones matrix for a general $q$-plate with retardance $\delta$ is given by
\begin{equation}
\mathcal{M}(\delta, \alpha(\phi))= 
\begin{bmatrix}
        \cos \frac{\delta}{2} + i\sin\frac{\delta}{2} \cos 2 \alpha(\phi) & i\sin\frac{\delta}{2} \sin 2 \alpha(\phi) \\[10pt]
        i\sin\frac{\delta}{2} \sin 2 \alpha(\phi) & \cos \frac{\delta}{2} - i\sin\frac{\delta}{2} \cos 2 \alpha(\phi)
\end{bmatrix}\in \text{SU(2)}.
\label{matrix01}
\end{equation}
This matrix features some special properties. This is a symmetric matrix, satisfying the condition $\mathcal{M}(\delta, \alpha(\phi)) = [\mathcal{M}(\delta, \alpha(\phi))]^{T}$. The Jones matrices corresponding to the $q^{Q}$- and $q^{H}$-plate satisfy the eighth and fourth roots of the identity element, respectively, expressed as 
\begin{equation}
\left[\mathcal{M}(\pi/2, \alpha(\phi))\right]^{8} = \mathbb{I} \quad \text{and} \quad \left[M(\pi, \alpha(\phi))\right]^{4} = \mathbb{I}.
\end{equation}
\noindent
The diagonal elements are complex conjugates, i.e., $\mathcal{M}_{11}(\delta, \alpha(\phi)) = [\mathcal{M}_{22}(\delta, \alpha(\phi))]^{*}$, while the off-diagonal elements are identical i.e., $\mathcal{M}_{12}(\delta, \alpha(\phi)) = \mathcal{M}_{21}(\delta, \alpha(\phi))$ and are purely imaginary. Further, this is an SU(2) matrix $(\texttt{det}[\mathcal{M}(\delta, \alpha(\phi))]=1$ and $\mathcal{M}(\delta, \alpha(\phi))[\mathcal{M}(\delta, \alpha(\phi))]^{\dagger}=\mathbb{I})$.
\section{Holonomic condition in polarization optics}
\label{section4}
There are three topological parameters: the topological index (PH index) of the beam, the order of the topological sphere, and the $q$-value of the topological element. For topologically protected structured polarization states, the holonomic constraint equation is given by
\begin{equation}
({S_1^{(\eta)}})^2+({S_2^{(\eta)}})^2+({S_3^{(\eta)}})^2=({S_0^{(\eta)}})^2.
\label{CE1}
\end{equation} 
For a holonomically constrained polarization transformation, the initial, final and the trajectory of points representing intermediate states all should lie on the same HOPS. The components $\psi_{R}^\ell$ and $\psi_{L}^\ell$ of the HOPS beam (Eq. (\ref{beam})) are the complex amplitudes expressed as 
\begin{align}
    \psi_{R}^\ell &= \langle{R_\ell}|\psi_{\ell}\rangle = \cos{\left(\frac{\pi}{4}-\chi^{(\eta)}\right)}e^{-i\gamma^{(\eta)}}, \\
    \psi_{L}^\ell &= \langle{L_\ell}|\psi_{\ell}\rangle = \sin{\left(\frac{\pi}{4}-\chi^{(\eta)}\right)}e^{i\gamma^{(\eta)}}.
\end{align}
When the input HOPS beam described by Eq. (\ref{beam}) passes through the $q$-plate, the resulting output HOPS beam can be written as
\begin{align}
|\psi^{'}_\ell\rangle &= \mathcal{M}(\delta, \alpha(\phi))|\psi_\ell\rangle \nonumber \\
 &= \psi_1 |R\rangle + \psi_2 |L\rangle,
\label{eqn_psi01}
\end{align}
where the complex components $\psi_1$ and $\psi_2$ are given by
\begin{align}
    \psi_{1} &= \left[i\sin{\left(\frac{\pi}{4}-\chi^{(\eta)}\right)}\sin{\frac{\delta}{2}}e^{-i\left(2\alpha_{0}-\gamma^{(\eta)}\right)}\right]e^{-i(2q-\ell)\phi} +\left[\cos{\left(\frac{\pi}{4}-\chi^{(\eta)}\right)}\cos{\frac{\delta}{2}}e^{-i\gamma^{(\eta)}} \right]e^{-i\ell\phi},
    \label{psi_1}
\end{align}
\begin{align}
    \psi_{2} & = \left[i\cos{\left(\frac{\pi}{4}-\chi^{(\eta)}\right)}\sin{\frac{\delta}{2}}e^{i\left(2\alpha_{0}-\gamma^{(\eta)}\right)}\right]e^{i(2q-\ell)\phi} + \left[\sin{\left(\frac{\pi}{4}-\chi^{(\eta)}\right)}\cos{\frac{\delta}{2}}e^{i\gamma^{(\eta)}} \right]e^{i\ell\phi}.
    \label{psi_2}
\end{align}

\begin{figure}[t]
\centering
\includegraphics[width=0.63\linewidth]{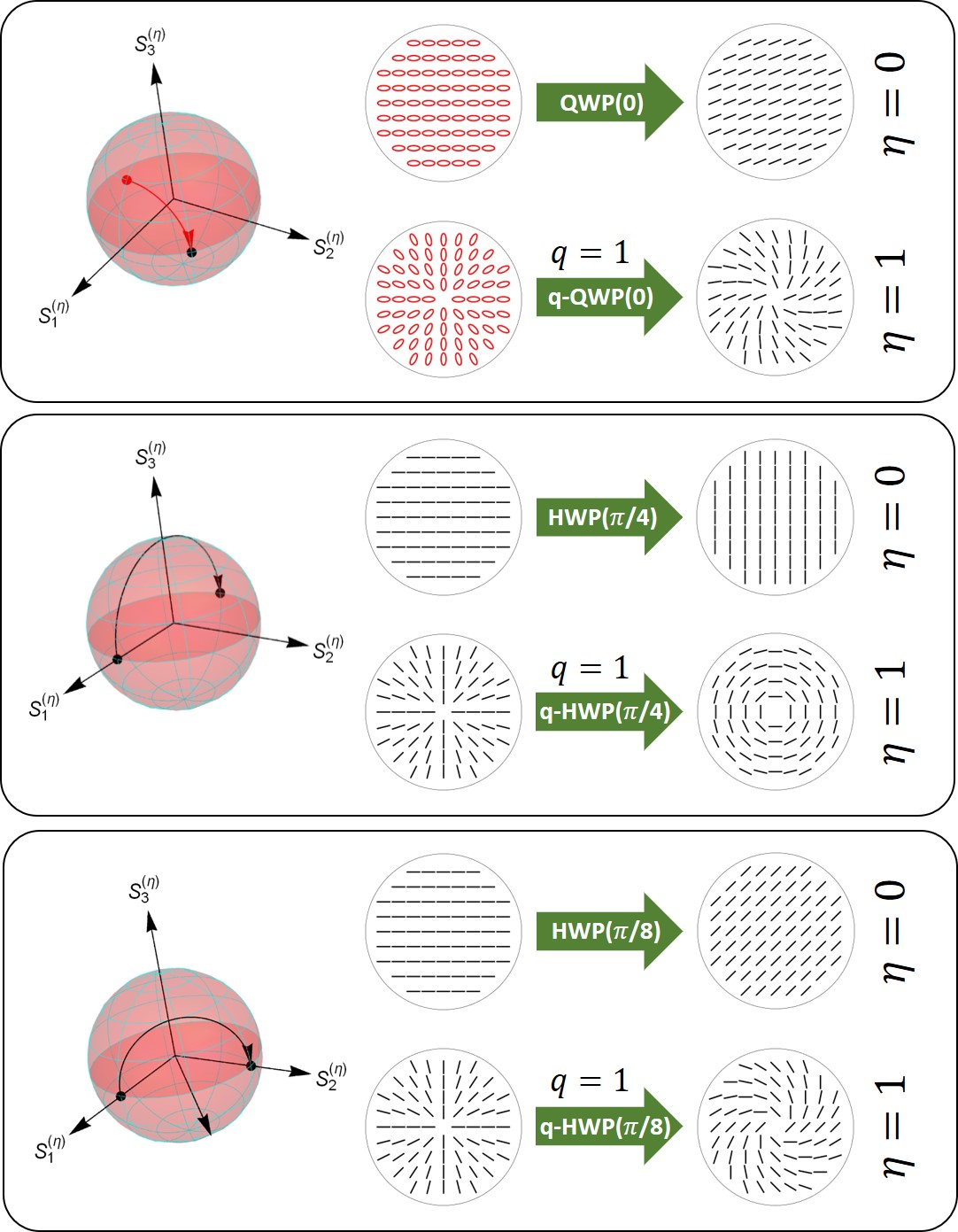}
\caption{(Color online). \textit{Holonomically constrained polarization transformations on HOPS of order $0$ and $1$. Polarization transformations are shown as rotations, with the rotation axis defined in the spheres. For HOPS ($\eta=1$) the rotation axis encloses an angle of $2\alpha_0$ with respect to $S_1^{(\eta)}$-axis and for HOPS ($\eta=0$), this angle is $2\alpha$ with the $S_1^{(0)}$-axis.}}
\label{q03}
\end{figure}

It is evident from the above equations that, for the output beam to remain on the same sphere, the condition $2q - \ell = \ell \Rightarrow q = \ell$ must be respected.
This implies that a holonomically constrained polarization transformation occurs when the three parameters, namely, $l$, which enters through Eq.~(\ref{beam}), the beam index $\eta$ (order of the HOPS) and the $q$-value of the plate, must all be equal, meaning that all topological features should match:
\begin{equation}
\label{holonomy}
\underbrace{l=\eta=q}_{\text{Holonomy condition}}
\end{equation}
With this condition, the output beam respects the form given in Eq. (\ref{beam}). Further, under this condition, the components $\psi_{1}$ and $\psi_{2}$ in Eq. (\ref{eqn_psi01}) is expressed as
\begin{align}
\psi_1 &= (i\xi_{1}s + \xi_{2}c)e^{-i\ell\phi}, \quad  \quad
\psi_2 = (i\xi_{2}s + \xi_{1}c)e^{i\ell\phi},
\label{holocomp}
\end{align}
where,
\begin{equation}
\xi_{1} = \sin\left(\frac{\pi}{4} - \chi^{(\eta)}\right), \quad
\xi_{2} = \cos\left(\frac{\pi}{4} - \chi^{(\eta)}\right), \quad
s = \sin\left(\frac{\delta}{2}\right) e^{i(2\alpha_0 - \gamma^{(\eta)})}, \quad
c = \cos\left(\frac{\delta}{2}\right) e^{i\gamma^{(\eta)}}.
\label{holocomp_01}
\end{equation}
\indent
The initial and final states are found to belong on the same HOPS, since the transformation satisfies Eq. (\ref{holonomy}). The trajectory that connects the initial and final states also lies on the surface of the HOPS by the same argument. For all retardation values of the $q$-plate, as long as the holonomic condition is respected, the transformation remains holonomic.\\
\indent
The trajectory between polarization states $A$ and $B$ on the HOPS can be traced by advancing the point that represents the input state iteratively through $N$ discrete steps. In each step, the point is advanced by an infinitesimal angular displacement $\Delta \delta$, such that the initial state $A = A_0$ and the next state $A_1$ are related by $A_1 = \mathcal{M}(\delta, \alpha(\phi)) A_0$. The \textit{j}-th state is given by $A_j = \mathcal{M}(\delta, \alpha(\phi)) A_{j-1}$, and $B = A_N$. Here, $\Delta \delta$ is determined by $\Delta \delta = \frac{\delta}{N}$, where $\delta$ is the total retardation produced by the $q$-plate, and $\mathcal{M}(\delta, \alpha(\phi))$ represents the SU(2) Jones matrix of the $q$-plate (Eq. (\ref{matrix01})).\\
\indent
Since all the points on a given HOPS represent beams with the same topological index, the topological index is conserved in any holonomically constrained polarization transformation. Polarization transformation by a retarder of charge $q$ with offset angle $\alpha_0$ on the polarization distribution with PH index $\eta$ can be seen as an SO(3) rotation \cite{yao2023quantitative, umar20252} on a HOPS of order $\eta = q$. The two points representing the final and initial states are connected by a circular arc, with the initial state rotated through an angle $\delta$ about a rotation axis lying in the equatorial plane of the sphere, oriented at an angle $2\alpha_0$ with respect to the reference axis ($S_{1}^{(\eta)}$-axis), as depicted in Fig. \ref{q03}. The purpose of Fig. \ref{q03} is to show some holonomic polarization transformation (and holonomic trajectory) on topological spheres of order $\eta = 0$ and $\eta = 1$, achieved using waveplates with topological charges $q = 0$ and $q = 1$, respectively. In this figure, the notation QWP/HWP($\alpha_{0}$) indicates a waveplate with $q = 0$, where $\alpha_0 = \alpha$ represents the fast-axis orientation. On the other hand, the notation $q$-QWP($\alpha_{0}$)/$q$-HWP($\alpha_{0}$) refers to a waveplate with $q \neq 0$, where $\alpha_0$ denotes the offset angle of the $q$-plate.
\section{Non-holonomic polarization transformation}
\label{section5}
\begin{figure}[h]
\centering
\includegraphics[width=0.55\linewidth]{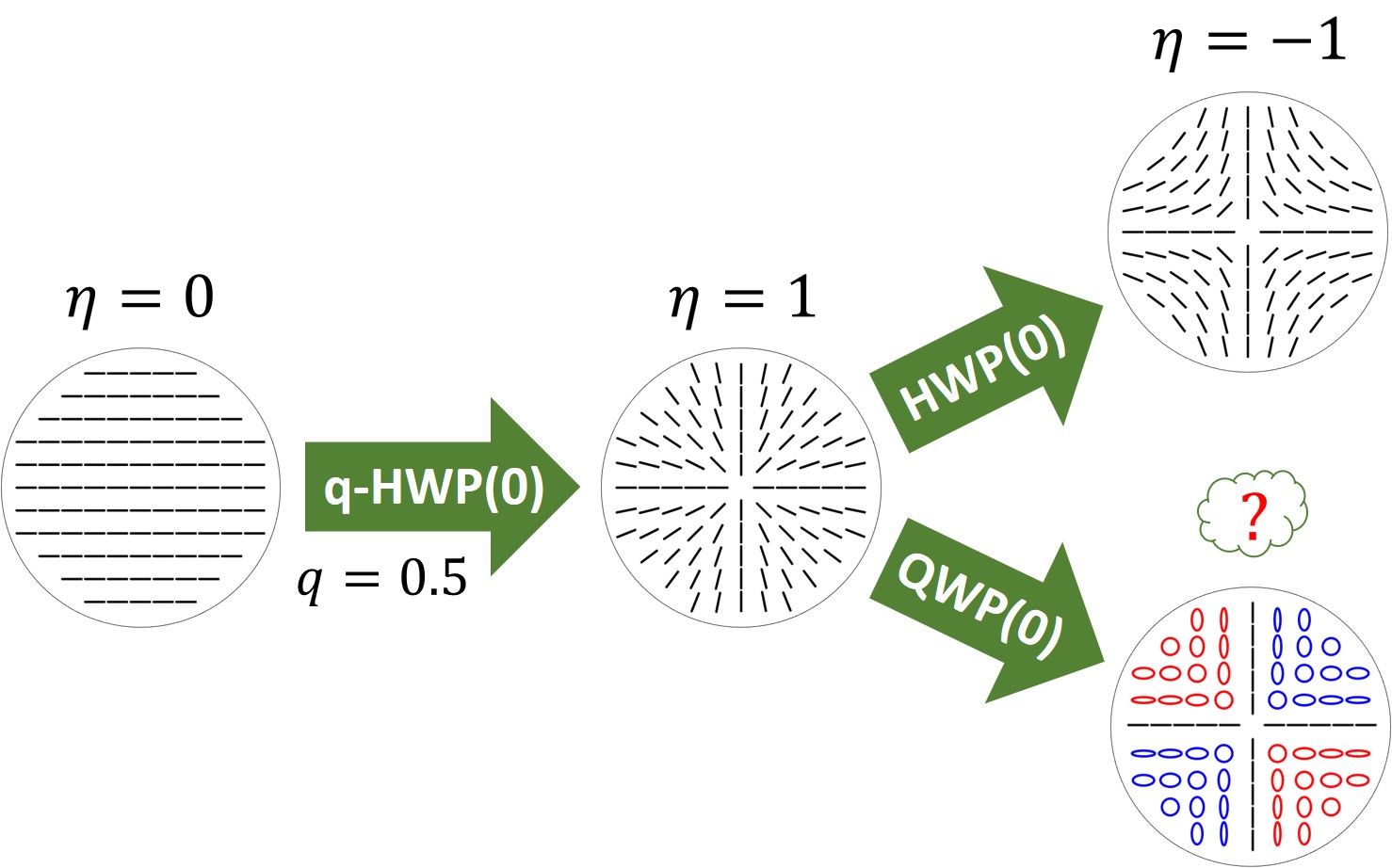}
\caption{(Color online). \textit{Illustration of the non-holonomic polarization transformations. The state of the light after passing through any of the elements that violates the holonomic condition $\eta=q$ results in beams that do not belong to the same sphere.}}
\label{q04}
\end{figure}

The polarization transformation is non-holonomic, when the condition given in Eq. (\ref{holonomy}) is not met. In such a case, a HOPS passing through a retarder will result in a polarization distribution that cannot be written in the form given by Eq. (\ref{beam}).
Fig. \ref{q04} shows non-holonomic polarization transformations. A horizontally polarized light represented by a point on standard PS - HOPS $(\eta=0)$ is taken to another sphere - HOPS ($\eta=1$) by the use of a $q^{H}$-plate (or $q$-HWP) with $q=1/2$. In successive passage through a HWP ($q=0$), this point on HOPS $(\eta=1)$, representing radial polarization singularity, goes to another sphere - HOPS $(\eta=-1)$, representing anti-radial polarization singularity. This is the case of polarization singularity index inversion \cite{pal2017polarization}.  But the use of QWP ($q=0$) on radial polarization takes the point on HOPS $(\eta=1)$ to nowhere. Therefore, these transformations are non-holonomic. A detailed study is needed for non-holonomic transformations. 
\section{Topological index spaces}
\label{section6}
Based on holonomic transformations, one can define polarization optics in different topological index spaces. Polarization optics in the topological index \textit{zero} space corresponds to the usual polarization optics, which pertains to homogeneously polarized beams, homogeneous elements (also the combination of homogeneous elements) and the standard PS. For example, the standard PS and corresponding SU(2) gadget (combination of two QWP and one HWP arranged in any order) \cite{simon1990minimal} also belongs to the topological index zero space. This includes all polarization optics that existed before the use of polarization singularities and $q$-plates. All the states and operators belong to this topological index space. Now, the advent of singular beams and $q$-plates allows us to define the topological index space for higher-order topological spheres, along with the corresponding structured elements.
\[
\text{Topological index ($\eta=q$) space} \Rightarrow  \left\{
  \begin{array}{l}
    \text{Topological sphere of order $\eta$} \\
    \underset{\text{Structured element of charge $q$}}{}
  \end{array}
\right.
\]
In each topological index space, there are two members: a topological sphere and the corresponding structured element as well as the combination of the structured element, which is capable of performing holonomic transformations on the sphere. The elements that do not belong to the same index space produce an altogether different effect on the states. In Fig. \ref{q04}, it is shown that an HWP that belongs to the $q=0$ world produces unexpected results on the $\eta \ne 0$ worlds. A point with a given latitude and longitude means different polarization distributions in different spheres. Each world or index space is governed by its unique sets of states and operators.
\begin{figure}[t]
\centering
\includegraphics[width=0.7\linewidth]{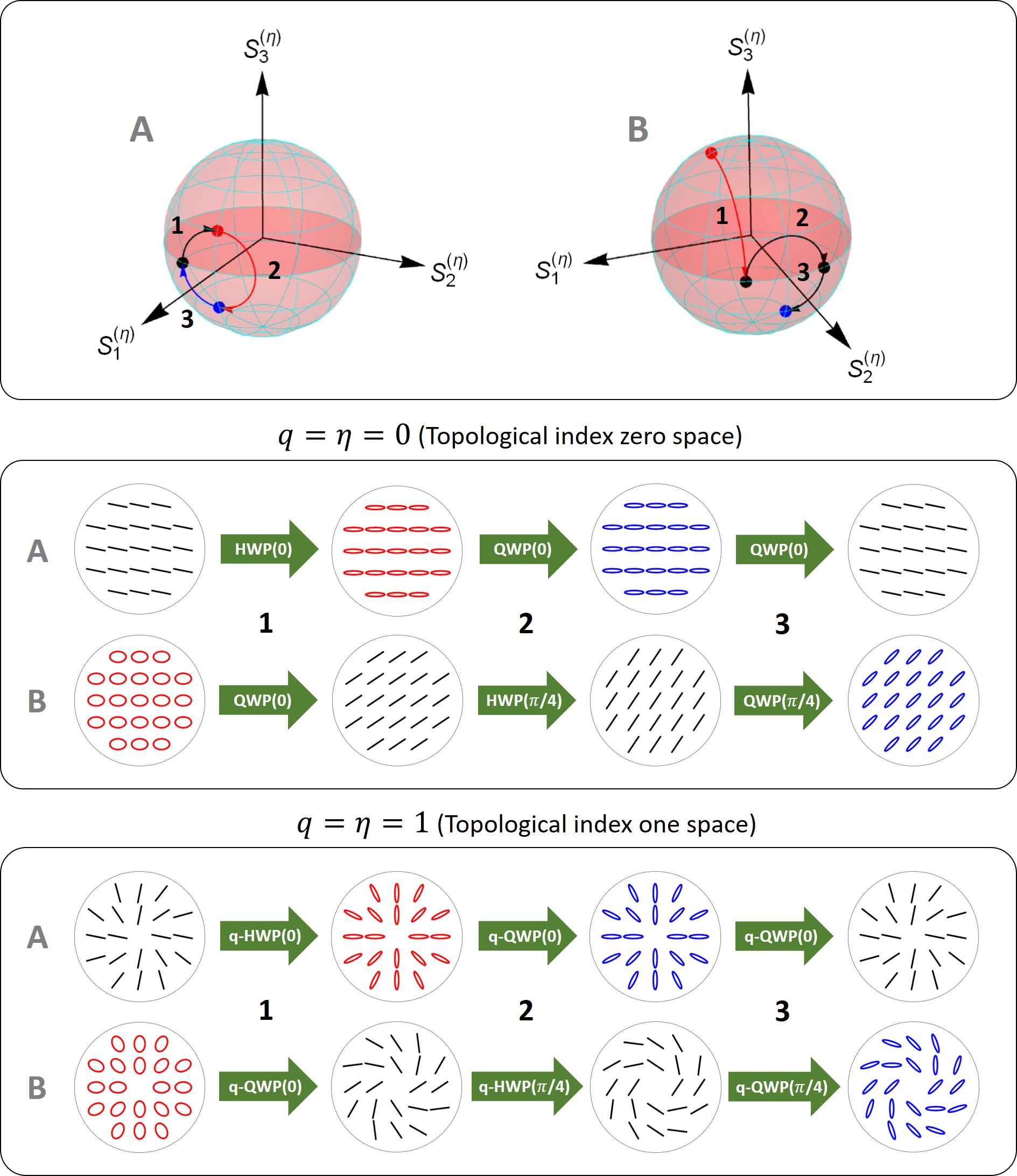}
\caption{(Color online). \textit{Polarization optics in topological index zero space and in topological index one space.}}
\label{q06}
\end{figure}
\section{Polarization optics in different index spaces}
\label{section7}
In Fig. \ref{q06}, polarization optics are shown in the space with the zero topological index and the space with the one topological index. Polarization states are transported from the initial state to two different states through the trajectories labeled $1$, $2$, and $3$ in two examples. In both configurations, light passes sequentially through a $q^{Q}$-plate, followed by a $q^{H}$-plate and then another $q^{Q}$-plate ($q^{Q}-q^{H}-q^{Q}$ polarization circuit). For the index zero space, where $\eta=q=0$, the polarization transformations corresponding to the two spheres are shown in the box just below the spheres, whereas for the index one space, where $\eta=q=1$, they are displayed in the bottom box. In the first sphere, all three polarization transformations share the same rotation axis. In contrast, in the second sphere, the rotation axis of the first transformation differs from those of the subsequent two, with each axis determined by the offset angle of the corresponding $q$-plates. It can be seen that in Fig. \ref{q06}, all the trajectories and initial, intermediate, and final states are constrained to the same HOPS as long as the conditions for holonomically constrained polarization transformations are met. All applicable rules hold within their respective spaces as long as holonomic transformations are performed.\\
\indent
The holonomic condition $\eta = q$ establishes that for a given topological sphere of order $\eta$, there exists a corresponding structured element, which enables the execution of a holonomic transformation on the HOPS. The concept of topological index space is built upon this relationship, where each family of topological index spaces consists of two components: one is the sphere, and the other is the corresponding waveplate. In Fig. \ref{q06}, the topological index zero space and index one space are presented for HOPS of order $\eta = 0$ and $1$, along with their corresponding polarization circuits $q^{Q}-q^{H}-q^{Q}$ with $q = 0$ and $q = 1$, respectively. Furthermore, for more instance, the polarization optics in the topological index two space and the topological index three space correspond to the HOPS of $\eta = 2$ with the $q$-plate of $q = 2$ and the HOPS of $\eta = 3$ with the $q$-plate of $q = 3$, respectively. Consequently, an infinite number of topological index spaces can be defined, as there are infinite HOPS, each with its corresponding structured element. Structured element that lie outside a particular topological index space produce inherently distinct transformations of polarization states.
\section{Conclusion}
\label{section8}
In summary, this paper introduces holonomic and non-holonomic transformations in the context of polarization optics. The conditions on the topological parameters of the beams, elements, and spheres required to achieve holonomically constrained polarization transformations are discussed in detail. A topological treatment of holonomic systems is needed, as many polarization transformations reported in the literature for beams with structured polarization are non-holonomic. A key outcome of this work is that these concepts enable the introduction of topological index spaces for polarization optics, providing a new framework for classifying and designing polarization transformations. Polarization optics admits an infinite number of topological index spaces. Elements that do not belong to a specific topological index space induce fundamentally different transformations on the polarization states.\\ 
\\
\\

\noindent
\textbf{Funding:} Science and Engineering Research Board (SERB) India (CRG/2022/001267).\\
\\
\noindent
\textbf{Disclosures:} The authors declare that there are no conflicts of interest.\\
\\
\noindent
\textbf{Acknowledgement:} MU gratefully acknowledges the fellowship support from IIT Delhi. MU also extends heartfelt thanks to the members of the Singular Optics Lab at IIT Delhi for their invaluable support and encouragement throughout the course of this research.\\
\\

\newpage
\bibliographystyle{elsarticle-num}
\bibliography{holonomy}

\end{document}